\documentclass[aps,preprint,prd,epsfig,nofootinbib]{revtex4}

\usepackage[utf8]{inputenc}
\usepackage[T1]{fontenc} 
\usepackage[dvipsnames]{xcolor}
\usepackage{multirow}
\usepackage{slashed}
\usepackage[normalem]{ulem}
\usepackage{sidecap}
\usepackage{mathtools}
\usepackage{caption}
\usepackage{subcaption}
\usepackage[toc]{appendix}
\usepackage{hyperref}
\usepackage{amsmath}
\usepackage{graphicx}
\usepackage{fancyhdr}
\usepackage{multirow}
\usepackage{booktabs}
\usepackage{makecell}
\usepackage{tabulary}
\usepackage{feynmp}

\newcommand{\be}{\begin{equation}}
\newcommand{\ee}{\end{equation}}
\newcommand{\bea}{\begin{eqnarray}}
\newcommand{\eea}{\end{eqnarray}}
\newcommand{\bes}{\begin{subequations}}
\newcommand{\ees}{\end{subequations}}

\newcommand{\bc}{\begin{center}}
\newcommand{\ec}{\end{center}}
\def\equationautorefname~#1\null{Eq.~(#1)\null}

\begin{document}

\title{Bounds on quark mixing, $M_{Z^{\prime}}$ and $Z-Z^{\prime}$ mixing angle  from  flavor changing neutral processes in a 3-3-1 model }

\author{Vin\'icius Oliveira$^a$}\email{vlbo@academico.ufpb.br
} 
\author{C. A. de S. Pires$^{a}$}\email{cpires@fisica.ufpb.br} 

\affiliation{$^{a}$Departamento de F\'isica, Universidade Federal da Para\'iba, Caixa Postal 5008, 58051-970, Jo\~ao Pessoa, PB, Brazil} 

\date{\today}

\begin{abstract}
Meson and anti-meson mixing processes constitute an important source of constraints on models that give tree level contributions to flavor violating neutral processes. In electroweak  $ \text{SU}(3)_\text{L} \times \text{U}(1)_N $  models, where anomaly cancellation requires that one family of quarks transforms differently from the other ones, processes involving flavor changing neutral currents gain  tree level contributions mediated by  gauge and scalar fields. Here, we firstly  investigate the   contributions of the neutral scalar that mimics the standard Higgs to the  $K^0-\bar K^0$, $B^0-\bar B^0$,  and  $D^0-\bar D^0$  mixing processes  and  confront our predictions  with experiments. The results will determine the quark mixing matrices $V^{u,d}_L$. In possession of this information we, next, evaluate the contributions of the neutral gauge bosons $Z^{\prime}$ and $Z$  to the $K^0-\bar K^0$, $B^0-\bar B^0$,  and  $D^0-\bar D^0$  processes. This realistic approach to meson transitions will provide  severe bounds  on $M_{ Z^{\prime}}$ and $Z-Z^{\prime}$ mixing angle.
\end{abstract}

\maketitle

\section{Introduction}
 Flavor changing neutral processes (FCNP) arise as a natural outcome of the electroweak $\text{SU}(3)_\text{L}\times \text{U}(1)_\text{N}$ models \cite{Singer:1980sw,Pisano:1992bxx,Frampton:1992wt,Foot:1994ym} because anomaly cancellations require at least that one  family of quark transforms differently from the others \cite{Pisano:1996ht,Frampton:1992wt}. These models predict FCNP mediated by  neutral gauge and  scalar fields \cite{Ng:1992st,Liu:1993gy,GomezDumm:1993oxo,Foot:1992rh} and, for example, the meson mixing  $K^0-\bar K^0$, $B^0-\bar B^0$, and  $D^0-\bar D^0$  processes   provide  an important source of constraint on the parameters of the models \cite{Ng:1992st,Liu:1993gy,GomezDumm:1993oxo,Foot:1992rh,Long:1999ij,Buras:2012dp,Crivellin:2013wna,Buras:2014yna,Crivellin:2015era,Queiroz:2016gif,CarcamoHernandez:2022fvl,Fan:2022dye,Alguero:2022est}.

Flavor changing neutral processes  mediated by $Z^{\prime}$ have received considerable attention within the $\text{SU}(3)_\text{C} \times \text{SU}(3)_\text{L} \times \text{U}(1)_\text{N}$ (3-3-1) models \cite{Ng:1992st,Liu:1993gy,GomezDumm:1993oxo,Long:1999ij,Rodriguez:2004mw,Cabarcas:2007my,Promberger:2007py,Benavides:2009cn,Cabarcas:2009vb,Cabarcas:2011hb,Jaramillo:2011qu,Cogollo:2012ek,Machado:2013jca,NguyenTuan:2020xls,Queiroz:2016gif,Huitu:2019kbm}, but there are few works exploring these processes mediated by neutral scalars \cite{Cogollo:2012ek,Okada:2016whh,Huitu:2019kbm}. In the 3-3-1 model with right-handed neutrinos (331RHN)  three neutral scalars and one pseudoscalar give  contributions at tree level to meson mixing \cite{Cogollo:2012ek,Okada:2016whh,Huitu:2019kbm}. In order to address FCNP inside models with complex scalar sector, the first thing to do is to recognize, in the spectrum of the scalar of the model, the neutral scalar that plays the role of the standard Higgs. We then calculate  the Higgs contributions  to  $K^0 -\bar K^0$, $B^0 -\bar B^0$, and $D^0 -\bar D^0$ mixing with the aim of determining  the unitary matrices, $V^{u,d}_L$,  that mix the standard quarks. This is important because with such matrices in hand we can realistically probe the contributions of the 3-3-1 models to FCNP. This approach gives us the real power of FCNP in constraining  the parameters of the 3-3-1 models. 

With this in mind, we focus on the 331RHN and recognize the Higgs in the spectrum of scalar of the model  and then  obtain its Yukawa interactions with the standard physical quarks that lead to FCNP. Next we calculate the contributions of the Higgs to the mass differences associated with $K^0 -\bar K^0$, $B^0 -\bar B^0$, and $D^0 -\bar D^0$  transitions. We confront our predictions with  experiments and, as result, we determine $V^{u,d}_L$. In possession of $V^{u,d}_L$, we calculate the contributions of $Z^{\prime}$ to $K^0 -\bar K^0$, $B^0 -\bar B^0$, and $D^0 -\bar D^0$ and derive  bounds on $M_{Z^{\prime}}$. Finally, we calculate the contribution of the standard neutral gauge boson to such processes and obtain  upper bounds on the $Z-Z^{\prime}$ mixing angle.

This work is  organized as follows. In \autoref{sec.:II} we diagonalize the mass matrix of the CP-even scalars, recognize the standard Higgs and obtain their Yukawa interactions. In \autoref{sec:III} we calculate the contributions of the Higgs, $Z^{\prime}$ and of the standard neutral gauge boson $Z$ to the $\Delta m_{K,B,D}$ and obtain the bounds. In \autoref{sec:IV} we summarize our results and present our conclusions.

\section{Higgs-quarks  Yukawa interactions} \label{sec.:II}
We  restrict our investigation to the case where the first family of quarks transforms as triplet while the other two transform as anti-triplet. For the fermion representation and content we refer the reader to Ref. \cite{Long:1996rfd}.

In the 331RHN  the Yukawa interactions  involving  quarks and scalars  are composed by the following terms
\begin{eqnarray}
&-&{\cal L}^Q_Y \supset  g_{1a}\bar Q_{1_L}\rho d_{a_R}+ g_{ia}\bar Q_{i_L}\eta^* d_{a_R} \nonumber \\
&&+h_{1a} \bar Q_{1_L}\eta u_{a_R} +h_{ia}\bar Q_{i_L}\rho^* u_{a_R} + \mbox{H.c.}\,,
\label{yukawa1}
\end{eqnarray}
where $Q_{1_L}$ is the first family of quarks (triplet)  and $Q_{i_L}$  with $i=2,3$ are the second and third ones (anti-triplets)\footnote{For the most general set of Yukawa interactions allowed by the three triplet of scalars and the quarks, see Ref. \cite{Doff:2006rt}. All the other terms that appear there mix the standard quarks with the exotic ones. We avoid these terms because they violate lepton number explicitly.}.

The triplet of scalars  $\eta$ and $\rho$, together with $\chi$, form the original scalar content of the model. The scalar potential, the shift of the neutral components of the scalar fields that  develop VEV and the set of constraint equations that guarantee the model develops a global minimum  are found in Refs.\cite{deSPires:2007wat,Oliveira:2022vjo}.

We are interested exclusively in the  CP-even scalars. More specifically we want to recognize the CP-even scalar that plays the role of the standard Higgs. Then, according to Ref.\cite{deSPires:2007wat}  and considering the basis $(R_{\chi^{\prime}}, R_\eta,R_\rho)$, these scalars compose the following mass matrix , 
\begin{equation}
M_R^2=
\begin{pmatrix}
\lambda_1 v^2_{\chi^{\prime}}+fv_\eta v_\rho/4v_{\chi^{\prime}} & \lambda_4 v_{\chi^{\prime}}  v_\eta/2- f v_\rho/4 &  \lambda_5 v_{\chi^{\prime}}  v_\rho/2- f v_\eta/4\\
\lambda_4 v_{\chi^{\prime}}  v_\eta/2- f v_\rho/4 & \lambda_2 v^2_\eta+ fv_{\chi^{\prime}} v_\rho/4v_\eta & \lambda_6 v_\eta v_\rho/2-f v_{\chi^{\prime}}/4\\
\lambda_5 v_{\chi^{\prime}}  v_\rho/2- f v_\eta/4 &  \lambda_6 v_\eta v_\rho/2-f v_{\chi^{\prime}}/4 &  \lambda_3 v^2_\rho+ fv_{\chi^{\prime}} v_\eta/4v_\rho
\end{pmatrix}.
\label{Mu}
\end{equation}

It is quite impossible to diagonalize such matrix analytically. We then resort to the alignment limit characterized by the conditions , $\lambda_4=\lambda_5=\lambda$,  $v_\eta=v_\rho=v$ and $f=2\lambda v_{\chi^{\prime}}$, where $v = 175$ GeV represents the standard VEV. With this approach  $M_R^2$ get simplified
\begin{equation}
M_R^2=
\begin{pmatrix}
\lambda_1 v^2_{\chi^{\prime}}+\frac{1}{2}\lambda v^2 & 0 &  0\\
0& \lambda_2 v^2+ \lambda v_{\chi^{\prime}}^2/2 & \lambda_6 v^2 /2- \frac{1}{2}\lambda v_{\chi^{\prime}}^2\\
0 &  \lambda_6 v^2 /2- \frac{1}{2}\lambda v_{\chi^{\prime}}^2 &  \lambda_3 v^2+\lambda v_{\chi^{\prime}}^2/2
\end{pmatrix}.
\label{Mu}
\end{equation}

Diagonalizing analytically this matrix we obtain  $H=R_{\chi^{\prime}}$ whose mass is $m^2_H\approx \lambda_1 v^2_{\chi^{\prime}}+\frac{1}{2}\lambda v^2 $. The other eigenvalues are  $ m^2_{h_1}\approx \frac{1}{2}(\lambda_2+\lambda_3+\lambda_6)v^2 $, and $ m^2_{h_2}\approx m^2_{h_1}-\lambda_6 v^2+\lambda_1 v^2_{\chi^{\prime}}$ where $ h_1 =\frac{1}{\sqrt{2}}( R_\eta + R_\rho)$, and  $ h_2 = \frac{1}{\sqrt{2}}( R_\rho - R_\eta)  $. We soon recognize that $h_1$ must play the role of the standard Higgs. The other two neutral scalars, $H$ and $h_2$, are heavy particles with  their masses scaling with $v_{\chi^{\prime}}$. In view of this, it is reasonable to assume that the main scalar contribution to the meson transitions is given by $h_1$.  We follow this approach and, then, obtain the Yukawa interactions among $h_1$ and the quarks.

Let us consider the standard up quarks. For the basis $u=(u_1\,,\,u_2\,,\, u_3)$  the Yukawa interactions above provide the following  mass matrix
\begin{equation}
M_u=\frac{v}{\sqrt{2}}
\begin{pmatrix}
h_{11} & h_{12} & h_{13} \\
-h_{21} & -h_{22}  & -h_{23}\\
-h_{31} & -h_{32} & -h_{33}
\end{pmatrix},
\label{Mu}
\end{equation}
diagonalizing this matrix by a bi-unitary transformation
\begin{equation}
V_L^{u \dagger}M_u V_R^u=
\begin{pmatrix}
m_u & 0 & 0 \\
0 & m_c  & 0\\
0 & 0 & m_t
\end{pmatrix},
\label{MuD}
\end{equation}
we get the masses of the up quarks. The relation among the basis is given by
\begin{equation}
   \hat u_{L,R}=V^{\dagger u}_{L,R} u_{L,R}\,,
\end{equation}
with $\hat u=(u\,\,,\,\,c\,\,,\,\, t)^T$.

For the down quarks we  have the mass matrix
\begin{equation}
M_d=\frac{v}{\sqrt{2}}
\begin{pmatrix}
g_{11} & g_{12} & g_{13} \\
g_{21} & g_{22}  & g_{23}\\
g_{31} & g_{32} & g_{33}
\end{pmatrix},
\label{Md}
\end{equation}
diagonalizing this matrix by a bi-unitary transformation
\begin{equation}
V_L^{d \dagger}M_d V_R^d=
\begin{pmatrix}
m_d & 0 & 0 \\
0 & m_s  & 0\\
0 & 0 & m_b
\end{pmatrix},
\label{MdD}
\end{equation}
we get the masses of the down quarks. The relation among the basis is given by
\begin{equation}
   \hat d_{L,R}=V^{\dagger d}_{L,R} d_{L,R}\,,
\end{equation}
with $\hat d=(d\,\,,\,\,s\,\,,\,\, b)^T$.

Note that $V_{R}^{d,u}$ act on the quark singlets $u_R$ and $d_R$ which means there is no constraint on them. Consequently, without any loss for the quark phenomenology, we may assume $V_{R}^{d,u}=I$ . In this case Eqs.(\ref{Mu}), (\ref{MuD}), (\ref{Md}) and (\ref{MdD}) give
\begin{eqnarray}
   && g_{ia}=\sqrt{2}(V^d_L)_{ia}\frac{(m_{down})_a}{v}\,, \, g_{3a}=\sqrt{2}(V^d_L)_{3a}\frac{(m_{down})_a}{v}\,,\nonumber \\
   &&  h_{ia}=-\sqrt{2}(V^u_L)_{ia}\frac{(m_{up})_a}{v}\,, \, h_{1a}=\sqrt{2}(V^u_L)_{3a}\frac{(m_{up})_a}{v},
\end{eqnarray}
where $i=2,3$, $a=1,2,3$,  $(m_{down})_a=m_d, m_s,m_b$, and  $(m_{up})_a=m_u, m_c,m_t$.

After all this,  in the end of the day we obtain the following Yukawa interactions among $h_1$ and the physical standard quarks
\begin{eqnarray}
 {\cal L}_Y^{h_1}&=& \frac{h_1}{\sqrt{2}} \bar{\hat{u}}_{b_L} \left(  (V^u_L)_{1a} (V^u_L)_{b1}\frac{(m_{up})_a}{v}  + (V^u_L)_{ia} (V^u_L)_{bi}\frac{(m_{up})_a}{v} \right)\hat u_{a_R}  \nonumber \\
&+& \frac{h_1}{\sqrt{2}}\bar{ \hat{d}}_{b_L}\left(  (V^d_L)_{1a} (V^d_L)_{b1}\frac{(m_{dowm})_a}{v}+ (V^d_L)_{ia} (V^d_L)_{bi}\frac{(m_{dowm})_a}{v}\right)\hat d_{a_R}  + \mbox{H.c.}\,,
\label{YcaseII}
\end{eqnarray}
where the subscripts $a,b=1,2,3$ and $i=2,3$. 

Observe that the interactions in \autoref{YcaseII} lead inevitably to processes that violate  flavor mediated by $h_1$. In this regard, particularly important are its contributions to meson-antimeson mixing $K^0-\bar K^0$, $B^0-\bar B^0$ and $D^0-\bar D0$. We calculate such contributions with the aim of shedding light in the pattern of quark  mixing by determining the elements of  $V^{u,d}_L$ .

\section{Flavor Changing Neutral Current bounds on the parameters of the model}\label{sec:III}

\subsection{Higgs contributions}
As we can observe in the  \autoref{YcaseII}, the neutral meson-antimeson transitions $K^0 - \bar K^0$, $B^0 - \bar B^0$, and $D^0 - \bar D^0$ receive contributions from  $h_1$ by means of the following  effective interactions \cite{Gabbiani:1996hi} 
\begin{eqnarray}\label{lagrangean}
    &&{\cal L}^{eff}_{K}= \frac{1}{2 m_{h_1}^2} \left[ \bar{d}\left( C_K^R P_R - C_K^L P_L \right) s \right]^2      ,\nonumber \\ 
     &&{\cal L}^{eff}_{B}=   \frac{1}{2 m_{h_1}^2} \left[ \bar{b}\left( C_B^R P_R - C_B^L P_L \right) d \right]^2     ,  \\ 
     &&{\cal L}^{eff}_{D}=   \frac{1}{2 m_{h_1}^2} \left[ \bar{u}\left( C_D^R P_R - C_D^L P_L \right) c \right]^2       ,\nonumber 
\end{eqnarray}
where the coefficients $C^{L,R}_{K,B,D}$ are given by
\begin{eqnarray}
    &&C^R_K = \left[ \left( V_L^d \right)_{12}  \left( V_L^d \right)_{11}    + \left( V_L^d \right)_{i2}  \left( V_L^d \right)_{1i}  \right]    \frac{m_s}{v} \,, \nonumber \\
    &&C^L_K = \left[ \left( V_L^d \right)_{11}^*  \left( V_L^d \right)_{12}^*    + \left( V_L^d \right)_{1i}^*  \left( V_L^d \right)_{i2}^*  \right]    \frac{m_d}{v}         ,\nonumber \\
    &&C^R_B =  \left[ \left( V_L^d \right)_{11}  \left( V_L^d \right)_{31}    + \left( V_L^d \right)_{i1}  \left( V_L^d \right)_{3i}  \right]    \frac{m_d}{v}       \,, \\
    &&C^L_B = \left[ \left( V_L^d \right)_{31}^*  \left( V_L^d \right)_{11}^*    + \left( V_L^d \right)_{3i}^*  \left( V_L^d \right)_{i1}^*  \right]    \frac{m_b}{v}          \,, \nonumber \\
    &&C^R_D = \left[ \left( V_L^u \right)_{12}  \left( V_L^u \right)_{11}    + \left( V_L^u \right)_{i2}  \left( V_L^u \right)_{1i}  \right] \frac{m_c}{v}       \,, \nonumber \\
    &&C^L_D =   \left[ \left( V_L^u \right)_{11}^*  \left( V_L^u \right)_{12}^*    + \left( V_L^u \right)_{1i}^*  \left( V_L^u \right)_{i2}^*  \right] \frac{m_u}{v} \,, \nonumber
  \label{coeficient}
\end{eqnarray}
with $i=2,3$.

According to these interactions the contributions of $h_1$ to $\Delta m_K$, $\Delta m_B$ and $\Delta m_D$  are given by the expressions \cite{Gabbiani:1996hi}
\begin{equation}\label{eq: DeltamK_Higgs}
    \Delta m_K = \frac{m_K B_K f_K^2}{m_{h_1}^2}\left[\frac{5}{24} Re[ \left( C_K^L \right)^2 + \left( C_K^R \right)^2]\left(\frac{m_K}{m_s + m_d} \right)^2 + 2 Re[C_K^L C_K^R] \left( \frac{1}{24} + \frac{1}{4} \left(\frac{m_K}{m_s + m_d} \right)^2\right)\right],
\end{equation}
\begin{equation}\label{eq: DeltamB_Higgs}
    \Delta m_B =   \frac{m_B B_B f_B^2}{m_{h_1}^2}\left[\frac{5}{24} Re[ \left( C_B^L \right)^2 + \left( C_B^R \right)^2]\left(\frac{m_B}{m_d + m_b} \right)^2 + 2 Re[C_B^L C_B^R] \left( \frac{1}{24} + \frac{1}{4} \left(\frac{m_B}{m_d + m_b} \right)^2\right)\right]\,,
\end{equation}
\begin{equation}\label{eq: DeltamD_Higgs}
    \Delta m_D =  \frac{m_D B_D f_D^2}{m_{h_1}^2}\left[\frac{5}{24} Re[ \left( C_D^L \right)^2 + \left( C_D^R \right)^2]\left(\frac{m_D}{m_u + m_c} \right)^2 + 2 Re[C_D^L C_D^R] \left( \frac{1}{24} + \frac{1}{4} \left(\frac{m_D}{m_u + m_c} \right)^2\right)\right]\,,
\end{equation}
where $m_{K,B,D}$ are the masses of the mesons,  $B_{K,B,D}$ are the bag parameters and $f_{K,B,D}$ the decay constants. Observe that the only free parameters involved in these expressions are the elements of the mixing matrix  $(V^{u,d}_L)_{ij}$ which means that these contributions, in conjunction with the unitary constraint on $V^{ u,d}_L$ and the fact that $V_L^uV^{\dagger d}_L=V_{CKM}$,  will provide the pattern of mixing among the quarks by determining the elements of $V^{u,d}_L$. As we are going to see, this is very important in determining the contributions of other  neutral particles, as $Z^{\prime}$, to these transitions.    

\subsection{$Z^{\prime}$ contribution}

It is important to remember that, from the spectrum of 3-3-1 neutral particles that potentially  contribute to meson transitions, we are going to assume  that  $Z^{\prime}$ gives the main contribution. That said we have that in the 331RHN the effective lagrangians that characterizes the  contributions to $\Delta m_K$, $\Delta m_B$, and $\Delta m_D$ mediated by $Z^{\prime}$ are given by \cite{GomezDumm:1993oxo,Long:1999ij,Benavides:2009cn}
\begin{equation}
    {\cal L}^{eff (Z^\prime)}_{K}= \frac{4 G_F c_W^4}{3 \sqrt{2} \left( 3 - 4 s_W^2 \right)}\frac{M^2_Z}{M_{Z^\prime}^2} \left[\left(V_{L}^d\right)_{11}^* \left(V_{L}^d\right)_{12}  \right]^2|\bar d_L \gamma^\mu s_L|^2\,,
\end{equation}
\begin{equation}
    {\cal L}^{eff (Z^\prime)}_{B}= \frac{4 G_F c_W^4}{3 \sqrt{2} \left( 3 - 4 s_W^2 \right)}\frac{M^2_Z}{M_{Z^\prime}^2} \left[\left(V_{L}^d\right)_{11}^* \left(V_{L}^d\right)_{13}  \right]^2|\bar d_L \gamma^\mu b_L|^2\,,
\end{equation}
\begin{equation}
    {\cal L}^{eff (Z^\prime)}_{D}= \frac{4 G_F c_W^4}{3 \sqrt{2} \left( 3 - 4 s_W^2 \right)}\frac{M^2_Z}{M_{Z^\prime}^2} \left[\left(V_{L}^u\right)_{11}^* \left(V_{L}^u\right)_{12}  \right]^2|\bar u_L \gamma^\mu c_L|^2\,,
\end{equation}
which leads to 
\begin{eqnarray}\label{eq:Delta_m_k_Zp}
 \left(\Delta m_K \right)_{Z^\prime} = \frac{4 G_F c_W^4}{3 \sqrt{2} \left( 3 - 4 s_W^2 \right)} \left[\left(V_{L}^d\right)_{11}^* \left(V_{L}^d\right)_{12}  \right]^2 \frac{M^2_Z}{M_{Z^\prime}^2} f^2_K B_K m_K\,,
\end{eqnarray}
\begin{eqnarray}\label{eq:Delta_m_B_Zp}
 \left(\Delta m_B \right)_{Z^\prime} = \frac{4 G_F c_W^4}{3 \sqrt{2} \left( 3 - 4 s_W^2 \right)} \left[\left(V_{L}^d\right)_{11}^* \left(V_{L}^d\right)_{13}  \right]^2 \frac{M^2_Z}{M_{Z^\prime}^2} f^2_B B_B m_B\,,
\end{eqnarray}
\begin{eqnarray}\label{eq:Delta_m_D_Zp}
 \left(\Delta m_D \right)_{Z^\prime} = \frac{4 G_F c_W^4}{3 \sqrt{2} \left( 3 - 4 s_W^2 \right)} \left[\left(V_{L}^u\right)_{11}^* \left(V_{L}^u\right)_{12}  \right]^2 \frac{M^2_Z}{M_{Z^\prime}^2} f^2_D B_D m_D\,,
\end{eqnarray}
where $G_F$ is the Fermi constant, $M_Z$ is the mass of standard $Z$ boson, $M_{Z^{\prime}}$ is the mass of $Z^{\prime}$ and $\theta_W$ is the Weinberg angle. Except by $M_{Z^{\prime}}$, the values of these parameters are found in Ref. \cite{Workman:2022ynf}. Observe here that Eqs. (\ref{eq:Delta_m_k_Zp}), (\ref{eq:Delta_m_B_Zp}), (\ref{eq:Delta_m_D_Zp}) involve six free parameters, namely $(V^d_L)_{11}\,,\, (V^d_L)_{12}\,,\, (V^d_L)_{13}\,,\, (V^u_L)_{11}\,,\, (V^u_L)_{12}$ and $M_{Z^{\prime}}$. Thus it is imperative to know the elements of $V^{u,d}_L$ to extract any bound on $M_{Z^{\prime}}$. In previous work on this subject the elements of $V^{u,d}_L$ (the pattern of mixing) were postulated and then bounds on $M_{Z^{\prime}}$ were obtained, see Refs. \cite{Okada:2016whh,Cogollo:2012ek,Queiroz:2016gif}. Here we do the following: we obtain the elements of $V^{u,d}_L$ from  $h_1$ contributions to the meson transitions and use them in the expressions above to extract bound on $M_{Z^{\prime}}$. Besides of being possible, we think that  this approach is much more realistic than the previous one.

\subsection{$Z $ contribution  }
In 3-3-1 models $Z$ mix with  $Z^{\prime}$ to form  $Z_1$ and $Z_2$. Due to this mixing $Z_1$ starts to contribute to meson transitions, too. As we are going to see, such contributions are determined by the mixing angle $\phi$. We, then,  switch on the  $Z$-$Z^{\prime}$ mixing,  where now $Z_1= Z \cos \phi - Z^{\prime} \sin \phi$ play the role of the standard gauge boson. In this case $Z_1$ contribute to $K^0-\bar K^0$, $B^0-\bar B^0$, and $D^0-\bar D^0$ mass difference by means of the following effective lagrangian \cite{Long:1999ij},
\begin{equation}
    {\cal L}^{eff (Z)}_{K}= \frac{4 G_F c_W^4 \sin^2(\phi)}{3 \sqrt{2} \left( 3 - 4 s_W^2 \right)} \left[\left(V_{L}^d\right)_{11}^* \left(V_{L}^d\right)_{12}  \right]^2 |\bar d_L \gamma^\mu s_L|^2,
\end{equation}
\begin{equation}
    {\cal L}^{eff (Z)}_{B}= \frac{4 G_F c_W^4 \sin^2(\phi)}{3 \sqrt{2} \left( 3 - 4 s_W^2 \right)} \left[\left(V_{L}^d\right)_{11}^* \left(V_{L}^d\right)_{13}  \right]^2 |\bar d_L \gamma^\mu b_L|^2,
\end{equation}
\begin{equation}
    {\cal L}^{eff (Z)}_{D}= \frac{4 G_F c_W^4 \sin^2(\phi)}{3 \sqrt{2} \left( 3 - 4 s_W^2 \right)} \left[\left(V_{L}^u\right)_{11}^* \left(V_{L}^u\right)_{12}  \right]^2 |\bar u_L \gamma^\mu c_L|^2,
\end{equation}

which give the following contribution to $\Delta m_K$, $\Delta m_B$, and $\Delta m_D$
\begin{eqnarray}\label{eq:Z_phiK}
 \left(\Delta m_K \right)_{Z} = \frac{4 G_F c_W^4 \sin^2(\phi)}{3 \sqrt{2} \left( 3 - 4 s_W^2 \right)} \left[\left(V_{L}^d\right)_{11}^* \left(V_{L}^d\right)_{12}  \right]^2 f^2_K B_K m_K \,.
\end{eqnarray}
\begin{eqnarray}\label{eq:Z_phiB}
 \left(\Delta m_B \right)_{Z} = \frac{4 G_F c_W^4 \sin^2(\phi)}{3 \sqrt{2} \left( 3 - 4 s_W^2 \right)} \left[\left(V_{L}^d\right)_{11}^* \left(V_{L}^d\right)_{13}  \right]^2 f^2_B B_B m_B \,.
\end{eqnarray}
\begin{eqnarray}\label{eq:Z_phiD}
 \left(\Delta m_D \right)_{Z} = \frac{4 G_F c_W^4 \sin^2(\phi)}{3 \sqrt{2} \left( 3 - 4 s_W^2 \right)} \left[\left(V_{L}^u\right)_{11}^* \left(V_{L}^u\right)_{12}  \right]^2 f^2_D B_D m_D \,.
\end{eqnarray}
These contributions provide information exclusively on the mixing angle $\phi$. 

In summary, the Higgs contributions to the meson transitions cannot be neglected because they depend directly of $V^{u,d}_L$. As a result, such contributions  will provide information on $V^{u,d}_L$. In possession of this information, and assuming that the main contribution to the meson transitions from the  spectrum of 3-3-1 particles is due to $Z^{\prime}$, we extract bound on $m_{Z^{\prime}}$. From the contribution of $Z$ to the meson transitions we get information on the  mixing angle $\phi$. We do this analysis in the next section.

\section{Numerical analysis}
\subsection{Higgs contribution}
Observe that, according to the  coefficients $C^{R,L}_{K,B,D}$ in Eq. (\ref{coeficient}), the only free parameters in the expressions of the mass differences above are the elements of $V^{ u,d}_L$. Thus any bound on these mass differences fall exclusively on the elements of $V^{ u,d}_L$. It is in this point that lies the importance of the contributions of the standard Higgs to these processes, namely  assuming that such contributions recover the experimental errors, as consequence, we obtain a realistic proposal for the pattern of quark mixing  $V^{ u,d}_L$ \footnote{Of course that we have to have in mind that this is possible based in a certain set of reasonable  assumptions as, for example, that the right-handed quarks come in a diagonal basis.}. 

In general, neglecting CP phases, unitarity on $V^{u,d}_L$ imposes  these mixing matrices  be parameterized by three angles each. Moreover they lead to the CKM matrix by means of the relation
\begin{equation}
V^u_L V^{ d \dagger}_L=V_{\text{CKM}},
\label{ckmrelation1}
\end{equation}
which allows us to eliminate $V^u_L$ in favor of $V^d_L$ in the expression above
\begin{equation}
V^u_L =V_{\text{CKM}} V^d_L.
\label{ckmrelation2}
\end{equation}
In this way we have that, once $V_{\text{CKM}}$ is determined by experiments, in determining $V^d_L$ we automatically  obtain $V^u_L$.

Following the standard parametrization for a $3\times 3$ unitary matrix we write $V^d_L$  as  \cite{PhysRevLett.53.1802, Workman:2022ynf}
\begin{equation}
    V^d_L = 
\begin{pmatrix}
c_{12}c_{13}  &  s_{12}c_{13} &  s_{13}  \\
 -s_{12}c_{23} - c_{12} s_{23} s_{13}   & c_{12} c_{23} - s_{12}s_{23} s_{13}    &  s_{23}c_{13}\\
s_{12}s_{23}-c_{12}c_{23}s_{13}    &-c_{12}s_{23}-s_{12}c_{23}s_{13}     &c_{23}c_{13}  
\end{pmatrix},
\label{VdL_parametrization}
\end{equation}
where $s_{ij}=\sin \theta_{ij}$, $c_{ij}=\cos \theta_{ij}$. The parametrization of   $V^u_L$ in terms of $s_{ij}$ and  $c_{ij}$ is obtained substituting (\ref{VdL_parametrization}) in (\ref{ckmrelation2}). With $V^{u,d}_L$ parametrized by the angles $\theta_{ij}$ we substitute them in the expressions to the coefficients $C^{R,L}_{K,B,D}$ given above and  plug such coefficients in the expressions to $\Delta m_{K,B,D}$. After all this we are ready  to obtain constraints on  these three angles from Eqs. \ref{eq: DeltamK_Higgs}, \ref{eq: DeltamB_Higgs} and \ref{eq: DeltamD_Higgs}  by imposing that the Higgs contributions respect the experimental values of  $\Delta m_{K,B,D}$ . This is reasonable because we  have a system involving three equations, namely  \ref{eq: DeltamK_Higgs},  \ref{eq: DeltamB_Higgs}, \ref{eq: DeltamD_Higgs} and three variables ($\theta_{12}$, $\theta_{23}$ and $\theta_{13}$). However each equation of the system behaves likes $(\theta_{ij})^4$ which implies in many set of $(\theta_{13}\,,\,\theta_{23}\,,\, \theta_{12})$ as solution to the system. 

Besides the loop standard model contributions ($\Delta m_\text{SM}$)   to these mass differences present a good agreement with experiments \cite{Buchalla:1995vs,DiLuzio:2019jyq,DeBruyn:2022zhw},  we have that $\Delta m_\text{SM}$ still involves a considerable amount of uncertainty due to errors in  QCD corrections \cite{Wang:2019try,Wang:2022lfq}. We then find  reasonable to assume that all the contributions from new physics fall inside the experimental error. This is even more justifiable inside the 3-3-1 model because the model itself provides new  contributions at loop level to the mass differences that will increase the uncertainties.  

Our input parameters are shown in \autoref{Tab:Input}, and   $V_{\text{CKM}}$ is  given in the PDG \cite{Workman:2022ynf}
\begin{equation}
|V_{\text{CKM}}|=
\begin{pmatrix}
0.97435  & 0.22500 & 0.00369 \\
0.22486  & 0.97349 & 0.04182 \\
0.00857  & 0.04110 & 0.999118
\end{pmatrix}.
\label{CKM}
\end{equation}

\begin{table}
\begin{tabular}{ |p{3cm}|p{4.6cm}|p{4.5cm}|p{4.5cm}|  }
 \hline
 \multicolumn{4}{|c|}{Input Parameters} \\
 \hline
 Quark masses& $K^0$ &  $D^0$ & $B^0$\\
 \hline
  $m_u= 2.16 \ \text{MeV}$  &   $m_K= 497.611 \ \text{MeV}$  &  $m_D=1864.84 \ \text{MeV}$ &   $ m_B= 5279.66 \ \text{MeV}$\\
  $m_c= 1.27 \ \text{GeV}$  & $\sqrt{B_K} f_K= 131 \ \text{MeV}$  & $\sqrt{B_D} f_D= 212 \ \text{MeV}$   & $\sqrt{B_B}f_B= 210  \ \text{MeV}$\\
  $m_s= 93.4 \ \text{MeV}$  &  &  &  \\
  $m_b= 4.18 \ \text{GeV}$  &  &  &  \\
  $m_d = 4.67 \ \text{MeV}$ &  &  &  \\
 \hline
\end{tabular}
\caption{Input parameters \cite{Workman:2022ynf,10.1093/ptep/ptaa104,FlavourLatticeAveragingGroup:2019iem}.}
\label{Tab:Input}
\end{table}

The current experimental value for the mass difference are \cite{HFLAV:2019otj,PDBook,Chen:2021ftn,LHCb:2013zpr}
\begin{eqnarray} \nonumber
    &&\Delta m_K = \left(3.484 \pm 0.006 \right) \times 10^{-12} \text{ MeV}\,, \\
    &&\Delta m_D = \left(6.25316^{+2.69873}_{-2.8962} \right) \times 10^{-12} \text{ MeV}\,, \\ \nonumber
    &&\Delta m_B = \left(3.334 \pm 0.013 \right) \times 10^{-10} \text{ MeV}\,.
    \label{eq: Exp_DeltaM}
\end{eqnarray}

The experimental errors of the mass differences that we use to constrain new physics is,
\begin{eqnarray}\label{error}
 &&\Delta m_K= 0.006 \times 10^{-12}\mbox{ MeV},\nonumber \\
 &&\Delta m_D= 2.69 \times 10^{-12}\mbox{ MeV}, \\
 &&\Delta m_B= 0.013 \times 10^{-10}\mbox{ MeV}.\nonumber
\end{eqnarray}
That said ,  the idea here is to  demand that  any new physics contribution to these mass differences falls inside these  errors. The new contributions we are considering here are due to $h_1$, $Z$ and $Z^{\prime}$. Then the sum of such contributions must fill the error. In this point of the work we have to make another assumption regarding the amount of contribution of each mediator. We consider two scenarios. Without strong reason,  in one scenario we assume that $h_1$ get in charge of $10\%$ of the error while in the second scenario $h_1$ get in charge of $80\%$.  

For the case of the Higgs, whose interactions and contributions are presented above, after solving numerically  the set of Eqs. (\ref{eq: DeltamK_Higgs}), (\ref{eq: DeltamD_Higgs}) and (\ref{eq: DeltamB_Higgs})  by demanding that the Higgs provides a contribution to the mass differences exactly equal to the $80\%$, and $10\%$ of the errors in \autoref{error}, we obtain the following proposal of solutions to the mixing angles: 

For the scenario where $h_1$ contributes to for $80\%$ of the error, we get as solution, 
\begin{eqnarray}
 &&\theta_{12} = 1.580313\,,\nonumber \\
 &&\theta_{23} = 1.59713\,, \\
 &&\theta_{13} = -1.845500\,.\nonumber
\end{eqnarray}

These angles imply in the following pattern of quark mixing,
\begin{equation}
    V^d_L = 
\begin{pmatrix}
   0.0025814 &  -0.271249    & -0.96250    \\
  0.0171729  &   0.962379   &   -0.271168  \\
 0.999849   &  -0.015829    &0.00714249   
\end{pmatrix},
\label{VdL80}
\end{equation}
and by means of $V^u_L=V_{CKM}V^d_L$ we obtain,
\begin{equation}
    V^u_L = 
\begin{pmatrix}
    0.010068 &   -0.0478151    &-0.998804   \\
    0.0591118 &    0.875211   &-0.480109\\
   0.999695  &  0.0214141     &-0.0122575    
\end{pmatrix}
\label{VuL80}
\end{equation}

For the scenario where $h_1$ contributes to $10\%$ of the error, we get as solution,
\begin{eqnarray}
 &&\theta_{12} = 2.047993\,,\nonumber \\
 &&\theta_{23} = 0.978181\,, \\
 &&\theta_{13} = 1.555508\,.\nonumber
\end{eqnarray}

These angles imply in the following pattern of quark mixing,
\begin{equation}
    V^d_L = 
\begin{pmatrix}
   -0.0070215  & 0.0135799 & 0.999883  \\
    -0.115207 & -0.993261  & 0.0126809 \\
    0.993317 & -0.115105 & 0.00853869
\end{pmatrix},
\label{VdL10}
\end{equation}
and
\begin{equation}
    V^u_L = 
\begin{pmatrix}
    -0.0290977 &-0.210677  &0.977121  \\
-0.0721913 & -0.968689 & 0.237536 \\
  0.987645   &-0.15571& 0.0176213 
\end{pmatrix}.
\label{VuL10}
\end{equation}

These results are interesting  because they offers concrete patterns for the mixtures of the quarks. This allows us to probe concretely the  scale of 3-3-1 physics by means of the  $Z^{\prime}$ contribution to such mass differences. We do this in the next section. 

Before this, let us check if such patterns of mixing  are in agreement with ATLAS bounds on flavor changing neutral process involving the Higgs. We refer to  the ATLAS experimental  upper limit on the top quark decays $t \to h_1 c$ and $t \to h_1 u$ provided by the Yukawa interactions,
\begin{equation}
     {\cal L}^Y_{h_1} \supset \lambda_{tch_1} \Bar{t}h_1 c + \lambda_{tuh_1} \Bar{t}h_1 u + \text{H.C.}
\end{equation}
The ATLAS bounds on these top decays are found  in Ref. \cite{ATLAS:2017tas}  and impose  $ \lambda_{tch_1} < 0.13$ and $ \lambda_{tuh_1} < 0.13$. In this case we have to  verify if the Yukawa interactions in equation \autoref{YcaseII} respect such bounds. We have that for the two scenarios discussed above when $h_1$  contributes with $80\%$ of the error and when contributes with  $10\%$ of the error we get  $\lambda_{tch_1} \sim \mathcal{O} \left(  10^{-4}\right)$ and $\lambda_{tuh_1} \sim \mathcal{O} \left(  10^{-9}\right)$, respectively. 

\subsection{ $Z$ and $Z^{\prime}$ contributions}
Having in hand all this set of information  we are ready to obtain  the contributions of $Z$ and $Z^\prime$   to $\Delta m _{K,B,D}$ in both scenarios. In the first case, where we assumed that $h_1$ contributes with $80\%$ of the error, we are going to assume that $Z^{\prime}$ contributes with $10\%$  of the error, while in the second scenario, where we assumed that $h_1$ contributes with $10\%$ of the error, we are going to assume that $Z^{\prime}$ get in charge of $80\%$  of the error.  The remaining $10\%$, in the both scenarios, we assume is filled  with the contribution of $Z$.

\begin{figure}
    \centering
    \includegraphics[width=\columnwidth]{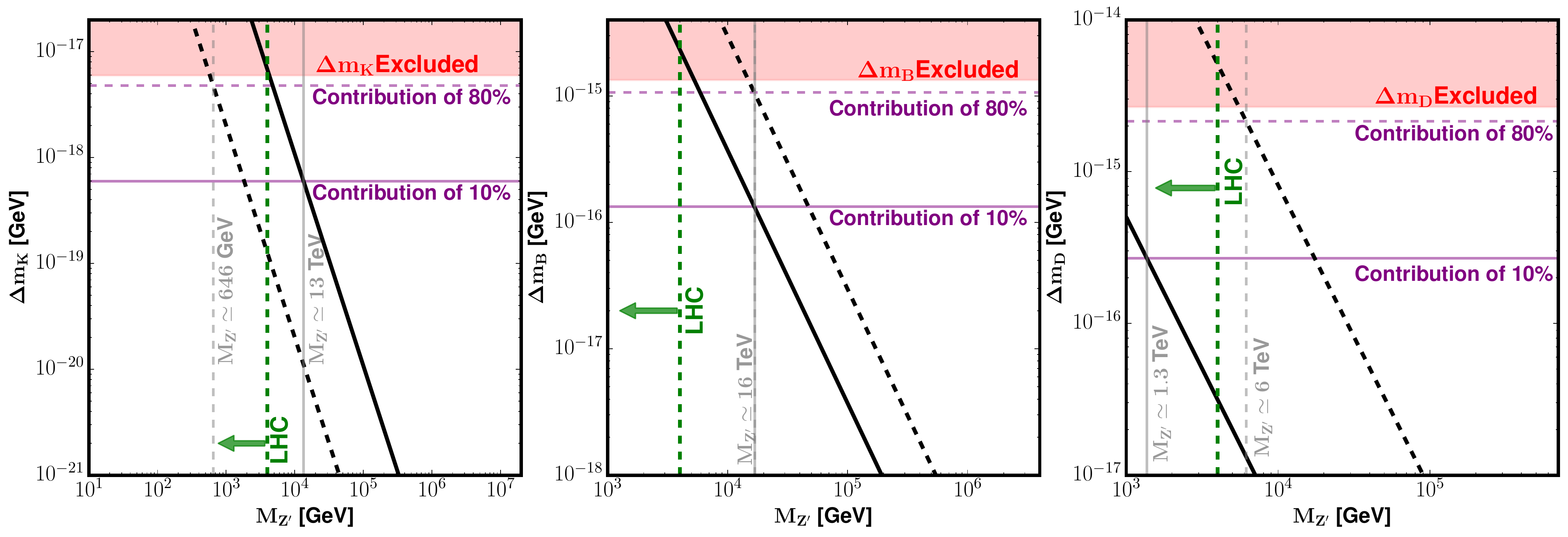}     
    \caption{Evolution of $\Delta m_K$, $\Delta m_B$, and $\Delta m_D$ in function of $M_{Z^\prime}$. The excluded red region represents the error of $\Delta m_{K,B,D}$.  The continuous (dashed) black lines represents the contribution of $Z^\prime$ for $80\%$ ($10\%$) of the error of difference of meson masses. The continuous (dashed) horizontal purple lines represents the contribution of $10\%$ ($80\%$) of $\Delta m_{K,B,D}$. }
  \label{fig:CaseI}
\end{figure}

In \autoref{fig:CaseI} we show our numerical results for the contributions of $Z^{\prime}$ to  $\Delta m _{K,B,D}$ in both cases. The excluded red region represents the error of $\Delta m _{K,B,D}$ as shown in \autoref{error}. The first panel show the $Z^{\prime}$ contribution for $\Delta m_K$ while the second and third panels represent its contributions for $\Delta m_B$ and $\Delta m_D$, respectively. The continuous and dashed black lines represent the evolution of $\Delta m_{K,B,D}$ in function of $M_{Z^\prime}$ as required by  Eqs. \ref{eq:Delta_m_k_Zp}, \ref{eq:Delta_m_B_Zp}, and \ref{eq:Delta_m_D_Zp}. 

In the three panels of \autoref{fig:CaseI}, the continuous black line represents the first case ($Z^\prime$ contributing with $10\%$ of error). The continuous purple horizontal line indicates the value of $10\%$ of $\Delta m_{K,B,D}$. The vertical green line represents the lower limit imposed by LHC \cite{Coutinho:2013lta}. In practical terms this means we are using the pattern quark mixing  given in Eqs. \ref{VdL80}, and \ref{VuL80}. As we can see in these pictures, the three transitions  are demanding a  heavy $Z^{\prime}$. Observe that $\Delta m_B$ put the strongest bound o $M_{Z^{\prime}}$. As we can see in the second panel, for $Z^{\prime}$ contribution constituting $10\%$ of the error of $\Delta m_B$ transition it demands $M_{Z^\prime} \approx 16$ TeV. This implies that the 3-3-1 model must belong to a scale energy around $v_{\chi^{\prime}} \approx 40 $TeV.

Concerning the second case, represented by the dashed black curves, where   $Z^\prime$ is supposed to contribute with $80\%$ of the error,  the strongest bound on $M_{Z^{\prime}}$ is also given by $\Delta m_B$ demanding the same  $M_{Z^\prime} \approx  16$ TeV since  the vertical dashed gray line in the second panel is upon the continuous. In these pictures the dashed purple horizontal lines indicate the value of $80\%$ of $\Delta m_{K,B,D}$ which means the dashed curves represent the pattern of mixing giving  in Eqs. \ref{VdL10}, and \ref{VuL10}.

In general we observe that, in both cases, with $Z^{\prime}$ contributing with $10\%$ or $80\%$ of the error, the bounds on $M_{ Z^{\prime}}$ constitute  the most severe bound on the 331RHN existent in the literature. We stress that this is due to the fact that we are using a pattern of quark  mixing allowed by the Higgs contributions to such processes. In Appendix A we display another  set of  possible patterns for $V^{u,d}_L$  allowed by the Higgs contributions and their respective higher  bound on $M_{Z^{\prime}}$ and $\phi$ from one of the mass difference $\Delta m_{K,B,D}$.  Observe also in these figures that such bound on $M_{Z^{\prime}}$ are not alleviated by varying the percent of the contribution of $Z^{\prime}$ to such transitions.   

\begin{figure}
    \centering
    \includegraphics[width=\columnwidth]{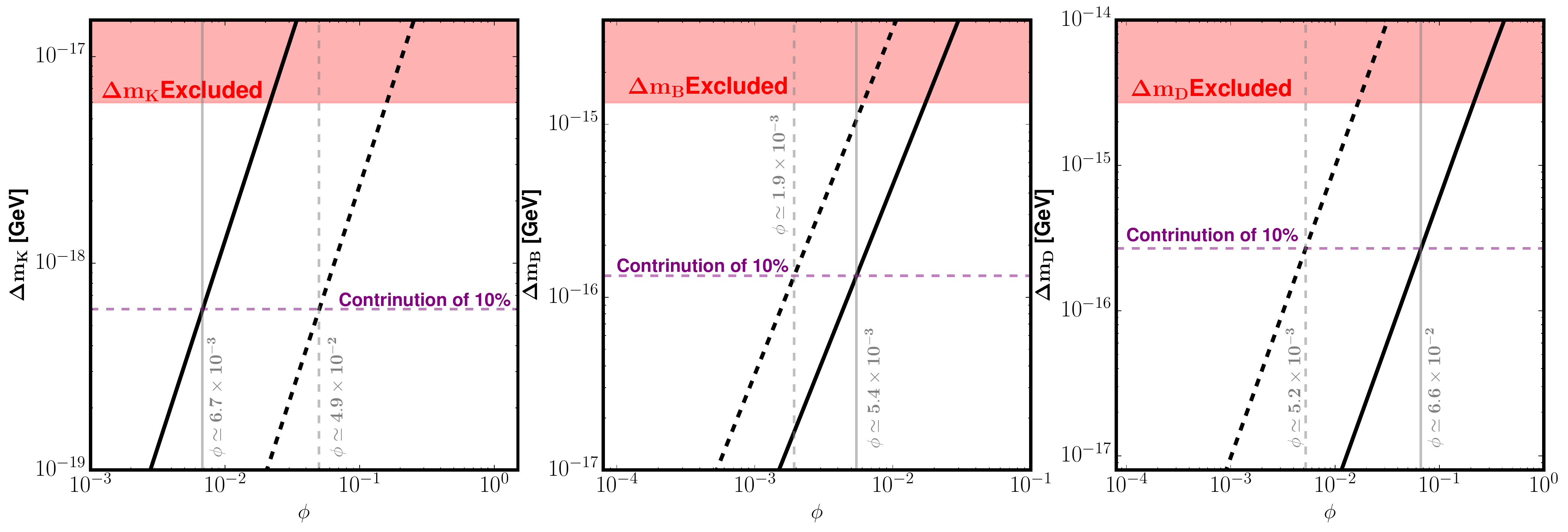}     
    \caption{Evolution of \autoref{eq:Z_phiK}, \autoref{eq:Z_phiB}, and \autoref{eq:Z_phiD} in function of $\phi$ (continuous black curves). }
  \label{fig:Z_phi}
\end{figure}

The results of the contributions of $Z$ for the remain $10\%$ of error is shown in \autoref{fig:Z_phi}, where the first panel represents the contribution of $Z$ to $\Delta m_K$ in function of $\phi$, and the second and third panels represents its contributions for $\Delta m_B$, and $\Delta m_D$, respectively. The excluded red region represents the error of $\Delta m _{K,B,D}$ as shown in \autoref{error}. The purple horizontal dashed lines indicates the value of $10\%$ of $\Delta m _{K,B,D}$, as we required. The black curves represents the evolution of \autoref{eq:Z_phiK}, \autoref{eq:Z_phiB}, and \autoref{eq:Z_phiD} in function of $\phi$. The continuous black curves means Higgs contributing with $80\%$ of the error, and the dashed black curves means Higgs contributing with $10\%$ of error. The contributions of $Z$ to these meson transitions translate exclusively in bounds on the $Z-Z^{\prime}$ angle mixing. As we can see in these pictures, the most stringent bounds are put by $\Delta m_K$ and $\Delta m_B$ where both require   $\phi \approx 6 \times 10^{-3}$. Such bounds  are close to the  previous ones found in the literature, see Refs. \cite{Long:1996rfd,Cogollo:2007qx}.   

We are aware that the bounds on $M_{Z^{\prime}}$ may change by taking $V^{u,d}_R \neq I$ since, in this case,  we may find space to handle  $V^{u,d}_L$ to a pattern that could result in  alleviation of the bound put here on $M_{Z^{\prime}}$ \footnote{Work in progress.}.

\section{Conclusions}\label{sec:IV}
In this work we considered the scalar sector of the 331RHN and obtained the mass matrix of the CP-even scalars. Applying the alignment limit we diagonalized it with the main aim of recognizing  the scalar that plays the role of the Higgs, $h_1$. Next, we obtained its Yukawa interactions with quarks in a scenario where the first family of quarks transforms as triplet. As expected, these interactions lead to flavor changing neutral processes. We then calculated the contributions of $h_1$ to the mass differences involved in meson-antimeson transitions. We assumed that any new contribution to the meson transitions cannot be larger than the experimental errors. Our investigation were divided in two scenario. In the one scenario $h_1$ get in charge of $80\%$ or the error and in the other scenario it get in charge of $10\%$. We stress that the unique variant parameters in the $h_1$ contributions are the quark mixing matrices $V^{u,d}_L$. The Higgs contributions serve solely to fix the elements of these mixing matrices. This is a very important result because with  $V^{u,d}_L$ in hand we may calculate the concrete contributions of the other neutral particles as $Z^{\prime}$ and $Z$ to the meson transitions. 

Throughout the paper we assumed that $h_1$, $Z$ and $Z^{\prime}$ give the dominant contributions to the error of the meson transitions which means the the sum of their contribution must practically fill the entire error. We then considered that $Z$ always get in charge of $10\%$ of the error. This means the sum of the contributions of  $Z^{\prime}$ and $h_1$ must sum $90\%$ of the error.  Whatever is the contribution of $Z^{\prime}$ to the error, it will provide  severe constrain on  the mass of $Z^{\prime}$ to values around $16$ Tevs.  The contribution of $10\%$ of $Z$ to the error translate in bounds on $Z-Z^{\prime} $ mixing that agree with previous ones.

Our study indicates that FCNC is an important source of constraint on the parameters of the 331RHN.  We remark that  the parameter space for new physics concerning meson mixing arises from the discrepancy between experimental and theoretical (standard model) predictions but such discrepancy is highly non-perturbative which leads to  considerable uncertainty. In view of our results, reducing this uncertainty is very important to the status of the 3-3-1 models. 

The results displayed in \autoref{fig:CaseI} and \autoref{fig:Z_phi} and in the appendix allow we conclude that FCNC in the form of meson transitions $K^0-\bar K^0$, $B^0-\bar B^0$ and $D^0-\bar D^0$ may be a very efficient way to constrain new physics  as is the case of the 331RHN. However, we stress that the weakness behind such analysis is the fact we do  not know yet the pattern of quark mixing matrices $V^{u,d}_L$. Determine such mixing from other sources of constraints, as Higgs and quarks  violating flavor decays,  will be decisive to the efficiency of meson transitions in constraining new physics.

\section*{Acknowledgments}
C.A.S.P  was supported by the CNPq research grants No. 	311936/2021-0 and V.O was supported by CAPES.

\appendix
\section{}\label{Ap.A}

\begin{tabular}{ |p{3cm}|p{3cm}|p{3cm}|p{3cm}|p{3cm}|  }
 \hline
 \multicolumn{5}{|c|}{Higgs Contribution of $80\%$} \\
 \hline
 $\theta_{12}$& $\theta_{23}$ & $\theta_{13}$ & $M_{Z^\prime}$ limit &$\phi$ limit\\
 \hline
 $1.6818$    &$-1.5212$ & $1.612$&  $31$ TeV &   $2 \times 10^{-3}$\\
  $-3.907867$ &$2.2188$ & $-1.58868$&  $87$ TeV &   $10^{-3}$\\
   $2.562462$&   $2.402145$  & $-1.586674$   &$90$ TeV &   $10^{-3}$\\
 $3.128632$   & $-1.86437$    & $0.002682$&   $248$ TeV &   $3.6 \times 10^{-4}$ \\
 \hline
\end{tabular}

\begin{tabular}{ |p{3cm}|p{3cm}|p{3cm}|p{3cm}|p{3cm}|  }
 \hline
 \multicolumn{5}{|c|}{Higgs Contribution of $10\%$} \\
 \hline
 $\theta_{12}$& $\theta_{23}$ & $\theta_{13}$ & $M_{Z^\prime}$ limit &$\phi$ limit\\
 \hline
   $1.737666$  &   $1.6224$  & $4.6562$ &  $22$ TeV  &   $4 \times 10^{-3}$\\
     $-1.138012$& $-1.280638$ & $-1.551463$  &  $19$ TeV  &   $4 \times 10^{-3}$\\
   $-1.025063$  &  $1.224614$   & $1.525326$ &  $56$ TeV  &   $1.6 \times 10^{-2}$\\
    $-0.000419$ &  $1.041301$   & $-0.003637$ &  $107$ TeV  &   $8\times 10^{-4}$\\
 \hline
\end{tabular}
\bibliography{biblio.bib}
\end{document}